\documentclass{article}

\pdfoutput=1 

\usepackage{arxiv}

\usepackage[utf8]{inputenc} % allow utf-8 input
\usepackage[T1]{fontenc}    % use 8-bit T1 fonts
\usepackage{hyperref}       % hyperlinks
\usepackage{url}            % simple URL typesetting
\usepackage{booktabs}       % professional-quality tables
\usepackage{amsfonts}       % blackboard math symbols
\usepackage{nicefrac}       % compact symbols for 1/2, etc.
\usepackage{microtype}      % microtypography
\usepackage{lipsum}		% Can be removed after putting your text content
\usepackage{graphicx}
\usepackage{natbib}
\usepackage{doi}
\usepackage{amsmath} % for math environments and more
\usepackage{booktabs}  % For formal tables
\usepackage{array}     % For table formatting
\usepackage{xcolor} % Include this package to use colors
\usepackage{appendix}  % Allows the appendices environment

\title{Leveraging Prompts in LLMs to Overcome Imbalances in Complex Educational Text Data}

\author{%
Jeanne McClure, Machi Shimmei, Noboru Matsuda, Shiyan Jiang \\
North Carolina State University, \\
Raleigh, NC 27695 \\
\texttt{\{jmmcclu3, mshimme, nmatsud, sjiang24\}@ncsu.edu}
}

\date{}

\renewcommand{\shorttitle}{\textit Leveraging Prompting}

\hypersetup{
pdftitle={A template for the arxiv style},
pdfsubject={q-bio.NC, q-bio.QM},
pdfauthor={David S.~Hippocampus, Elias D.~Striatum},
pdfkeywords={First keyword, Second keyword, More},
}

\begin{document}
\maketitle

\begin{abstract}

Background: The study addresses the challenge of imbalances in educational datasets, which is prominent in the education sector due to the varied cognitive engagement levels among students in their open responses. Traditional machine learning (ML) models often struggle with the complexity and nuanced nature of this data, leading to inadequate analyses, especially for minority data representations \citep{karimah2022automatic, radwan2017improving, yun2011effective}. Understanding students' cognitive engagement is vital as it reflects their mental investment in learning activities, which is closely linked to academic success \citep{fredricks2004school, blumenfeld2006motivation, corno1983role, pintrich2000role, schunk2014motivation}.

Objective: The objective of this paper is to investigate the efficacy of Large Language Models (LLMs) enhanced with assertions in tackling the complexities of imbalanced educational datasets, with a special focus on the precise classification of cognitive engagement levels from student texts. This exploration is underpinned by two critical research questions. The first seeks to evaluate how LLMs equipped with Prompt Engineering fare in comparison to conventional ML algorithms when dealing with the inherent challenges of imbalanced educational data. The second question delves into the specific contributions of integrating assertions into LLMs, examining how such augmentations can improve the models' effectiveness in handling the nuanced difficulties presented by imbalanced textual educational datasets. Through this inquiry, the study aims to shed light on the potential of LLMs and assertions in enhancing the accuracy and reliability of cognitive engagement classification, thereby addressing a significant gap in educational data analysis.

Methods: The study employed an 'Iterative - ICL PE Design Process' to compare traditional ML models against LLMs augmented with assertions (N=135). A sensitivity analysis on a subset (n=27) examined variance in model performance concerning classification metrics and cognitive engagement levels. This process involved the utilization of assertion-based prompt engineering, comparing the performance of traditional ML models to LLMs with assertions in classifying cognitive engagement from student texts in an educational setting \citep{shahriar2023assertion, brown2020language, wei2022emergent}.

Findings: LLMs with assertions significantly outperformed traditional ML models, especially in recognizing cognitive engagement levels with minority representation, showing up to a 32\% increase in F1-score. Incorporating targeted assertions into the LLM on the subset enhanced its performance by 11.94\%, primarily addressing errors from limitations in understanding context and resolving lexical ambiguities in student responses.

Implications: The study demonstrates the superior capability of LLMs, particularly when augmented with assertions, in addressing the nuanced challenges of imbalanced educational datasets. This advancement not only improves the accuracy of classifying cognitive engagement levels but also opens new avenues for data-driven educational research and practice. The findings suggest a potential paradigm shift towards employing advanced LLM techniques in educational settings to achieve a more nuanced and accurate analysis of student engagement, thereby enhancing learning outcomes. Future research should further explore the capabilities of LLMs across broader educational contexts and investigate additional methods to refine and expand their application in analyzing complex educational data \citep{shahriar2023assertion,zeng2023challenge}.
\end{abstract}

% keywords can be removed

\section{Introduction}
Understanding students' cognitive engagement (CE) at both the school and task levels is crucial, as it offers deep insights into their commitment to learning \citep{fredricks2004school}. This form of engagement, characterized by a student's deliberate and intentional approach to schoolwork and their willingness to invest the necessary effort in comprehending complex concepts and mastering challenging skills, serves as a key indicator of academic success \citep{fredricks2004school, blumenfeld2006motivation}. CE encompasses the psychological investment and effort driven by student motivation and strategies, alongside their dedication to learning \citep{corno1983role, fredricks2004school, pintrich2000role, schunk2014motivation}. 

While analyzing students' CE is crucial for enhancing learning experiences, a significant challenge arises from imbalanced datasets \citep{radwan2017improving}. These datasets often feature unevenly distributed categories and are typically small, not fitting the 'big data' criteria usually required for effective Machine Learning (ML) training. This size limitation, along with the disproportionate representation of majority and minority data, further complicates the training process in traditional analyses \citep{yun2011effective}. Traditional ML methods, commonly employed to classify CE, often struggle to adequately address these imbalances, raising concerns about the accuracy and reliability of their results. This issue presents a major hurdle in accurately assessing and interpreting CE, as the uneven representation of data can lead to skewed insights and potentially overlook critical aspects of student engagement \citep{karimah2022automatic}. This imbalance in datasets not only complicates the analysis but also raises concerns about the reliability and generalizability of the findings in diverse educational settings \citep{radwan2017improving}. 

The exploration of  LLMs provides a promising solution to the limitations of traditional ML approaches. Recent studies, including \citep{wu2021learning}, have highlighted the potential of prompt engineering in reducing the need for extensive training of case labeling which is imperative for imbalance data. LLMs employ techniques like In-context Learning (ICL) \citep{brown2020language} and Chain-of-Thought (COT) prompting \citep{wei2022chain}, enabling more nuanced and context-aware responses. ICL trains models using examples in specific contexts, improving with scaled model and corpus sizes, as seen in N-shot prompting \citep{brown2020language}. This is illustrated by Brown et al. \citeyearpar{brown2020language}’s few-shot learning, where LLMs process input-output pairs in-context, leading to better test-time predictions. Similarly, COT, by \citet{wei2022chain}, involves logical, step-by-step natural language reasoning. Furthering this, \citet{shahriar2023assertion} developed Assertion Enhanced Few-Shot Learning, incorporating domain-specific assertions in prompts to enhance accuracy and reduce errors. These innovations significantly boost LLMs' task-specific efficiency, surpassing traditional methods.

While LLMs have shown potential in educational research, their application has predominantly been refined to solve logical reasoning or arithmetic problems \citep{lee2024applying}, with limited exploration in addressing imbalanced datasets of education. Our study breaks new ground by applying LLMs with Prompt Engineering (PE) to this specific challenge. We hypothesize that LLMs, renowned for their nuanced language understanding, will surpass traditional ML algorithms in classifying cognitive engagement levels from student texts. Our exploration is guided by two research questions: RQ1 addresses the comparative efficacy of LLMs against traditional ML algorithms, and RQ2 investigates the role of assertions in overcoming contextual and lexical challenges within imbalanced datasets. Specifically:

\begin{enumerate}
    \item How do the results obtained from LLMs with PE compare to traditional Machine Learning algorithms in handling imbalanced educational data?
    \item In what ways does the integration of assertions enhance the efficacy of models when addressing the challenges associated with imbalanced textual educational datasets?
\end{enumerate}

This paper examines how AEFL mitigates issues  in imbalanced educational data analysis, revealing how these technologies can effectively address the challenges posed by uneven dataset distributions. By applying this cutting-edge technique, we uncover new possibilities for analyzing and interpreting complex educational data. Our findings demonstrate the advantage of AEFL in educational settings, especially where traditional ML methods fall short, opening new avenues for data-driven educational research and practice.

The rest of the paper is set up as follows: Section 2 delves into the background, highlighting the emergence of LLMs as a promising solution in education. Section 3 outlines our methodology, including the Iterative - ICL PE Design Process, and the experimental setup. The results and discussions are presented in Section 4, where we compare the performance of LLMs augmented with assertions against traditional ML models and discuss the impact of assertions on model efficacy and limitations. Finally, Section 5 concludes with our findings and future directions.

\section{Background}
The exploration of CE within educational research has significantly evolved, transitioning from a simplistic focus on student participation to a complex understanding of mental investment in learning activities. This shift is paramount for fully capturing the essence of engagement, as initially highlighted by \citet{craik1972levels} through their distinction between shallow and deep processing. Subsequent work by \citet{appleton2006measuring} and \citet{fredricks2004school} expanded the concept to encompass behavioral, emotional, and cognitive dimensions, underscoring engagement's multifaceted nature across various educational contexts. A pivotal insight from this exploration is the strong positive correlation between student learning and cognitive engagement, evidenced by \citet{chi2014icap}, which underscores the significant educational outcomes associated with deep cognitive processes.

CE distinguishes itself within the broader spectrum of educational engagement by focusing on the intensity of students' mental investment in learning. This stands in contrast to behavioral engagement's emphasis on participation and emotional engagement's concern with feelings towards learning \citet{blumenfeld2006motivation}. Such a distinction is crucial for educators and researchers dedicated to enhancing learning outcomes through targeted interventions.

Central to understanding and enhancing CE are theoretical frameworks and models like Bloom’s taxonomy, Corno and Mandinach’s model, and the ICAP model, as well as Wang et al.'s framework for connectivist learning contexts. These models provide comprehensive insights into the various dimensions and components of cognitive engagement, aiding researchers in designing effective studies, developing targeted interventions, and evaluating educational outcomes \citep{anderson2001taxonomy, bloom1956taxonomy, corno1983role, chi2014icap, chase2019teacher, hsiao2022developing, wang2016towards}.

Measuring CE, however, presents inherent challenges due to its complex and internal nature. As a latent construct, CE's assessment relies on inferences from behavioral indicators or through self-report measures \citep{chi2014icap, fredricks2004school, mccoach2013defining}. Traditional methods, including self-report questionnaires, surveys, and observational techniques, often inadequately capture the nuanced cognitive processes involved in learning. A variety of measures have been employed in past studies to gauge CE, such as self-reported scales, classroom observations, interviews, teacher ratings, experience sampling, eyetracking, physiological sensors, trace analysis, and content analysis \citep{greene2004predicting,smiley2011measuring, lee1993task, helme2001identifying, wigfield2008role, xie2019affordances, d2017advanced, bernacki2012effects, ireland2014language}. Nonetheless, the complexity of accurately assessing CE through these measures necessitates innovative approaches that more precisely reflect students' cognitive investment in their educational activities \citep{fredricks2004school}.

In educational research, traditional ML methods have extensively analyzed student data patterns but face limitations when addressing nuanced aspects like cognitive engagement. The problem is exacerbated by imbalanced datasets, leading to skewed insights and overlooking crucial engagement aspects, thus affecting the findings' accuracy, reliability, and generalizability across diverse educational contexts \citep{lee2012teacher, fredricks2004school}. This issue with imbalanced datasets, characterized by unevenly distributed categories and small sample sizes, highlights the need for specialized techniques to improve model performance and accuracy, ensuring a comprehensive understanding of CE across educational contexts \citep{chawla2010data, fernandez2018learning, kulkarni2020foundations, japkowicz2002class, bruce2020practical, lemaavztre2017imbalanced}.

The advent of LLMs presents a promising solution to the issues posed by imbalanced datasets in educational research. Recent breakthroughs in LLMs, particularly with ICL, COT and AEFL prompting techniques, have demonstrated their potential to generate nuanced, context-aware responses beyond the capabilities of traditional ML methods \citep{brown2020language, wei2022chain, shahriar2023assertion}. For example, \citet{savelka2023efficient} showcased how GPT-3.5 \& 4 could effectively classify student help requests in programming courses, illustrating the superior ability of LLMs to handle nuanced educational data. \citet{zeng2023challenge} delved into the cognitive and reasoning abilities of LLMs, highlighting the necessity for task-specific tuning to address complex reasoning challenges. \citet{cui2023divide} introduced the Divide-Conquer-Reasoning (DCR) framework to enhance the consistency and reliability of LLM-generated texts, vital for creating educational content. These examples reveal the capacity of LLMs to offer more accurate classification and analysis of CE, surpassing traditional ML methods in dealing with the intricacies of educational datasets. Additionally, \citet{lee2024applying} explored LLMs' use with CoT prompting to improve automatic scoring systems in science education, further indicating LLMs' potential to enhance the quality and reliability of educational content analysis.

By harnessing the intrinsic capacity of LLMs to interpret and utilize language within specific contexts, researchers can navigate the challenges posed by imbalanced datasets, facilitating a deeper understanding of student CE.

\section{Methodology}

\subsection{Context and Participants}
This study performs a secondary analysis on a dataset originally gathered to assess CE from student responses in a High School English Language Arts course's AI curriculum. The StoryQ curriculum \citep{chao2022storyq}, spanned three weeks with daily 45-minute classes, incorporated Machine Learning Practices through open-ended questions in eight modules but our analysis only evaluated three:  “Sentiment Analysis,” “Features and Models,” and “All Words.”
The initial study's diverse participant group of 28 students included 17 females, 7 males, and 4 non-specified gender individuals, spanning various grades and racial backgrounds. The racial composition was 43\% Black/African American, 17\% Hispanic/Latinx, 18\% White/Caucasian, with others choosing not to disclose. Students’ CE was evaluated using a modified Interactive-Constructive-Active- Passive (ICAP) framework by \citet{chi2014icap}, focusing on Constructive, Active, and Passive levels. Their open-ended responses (N = 840) were analyzed using the CE coding scheme, see Table~\ref{tab:ICAP}, yielding a Cohen’s kappa inter-rater reliability of 0.84.

\begin{table}[ht]
    \caption{ICAP: Cognitive Engagement Label Coding Scheme}
    \centering
    \begin{tabular}{clp{4cm}p{4cm}p{4cm}}
        \toprule
        Score & ICAP Level & Description & Indicator & Example \\
        \midrule
        2 & Constructive & New information is integrated with activated prior knowledge, and new knowledge is inferred. & Deep reasoning, synthesis of new ideas, or forming hypotheses. & ``I think that the model learned a large positive weight for the feature because if you came to an establishment then that would indicate that you did like it because you chose to come in the first place." \\
        \midrule
        1 & Active & Behaviors that cause-focused attention while manipulating. & Apply, Analyze, or Manipulating. & ``I think this gained a large amount of weight because it is a commonly used word." \\
        \midrule
        0 & Passive & Overt activities that are carried out mindlessly. & Recalling or Restating. & ``Amazing, clean, selection, try, regular, seating" \\
        \bottomrule
    \end{tabular}
    \label{tab:ICAP}
\end{table}

\subsection{Prompt Engineering Design}\label{sec:prompt_engineering}
\begin{figure}
    \centering
    \includegraphics[width=1\linewidth]{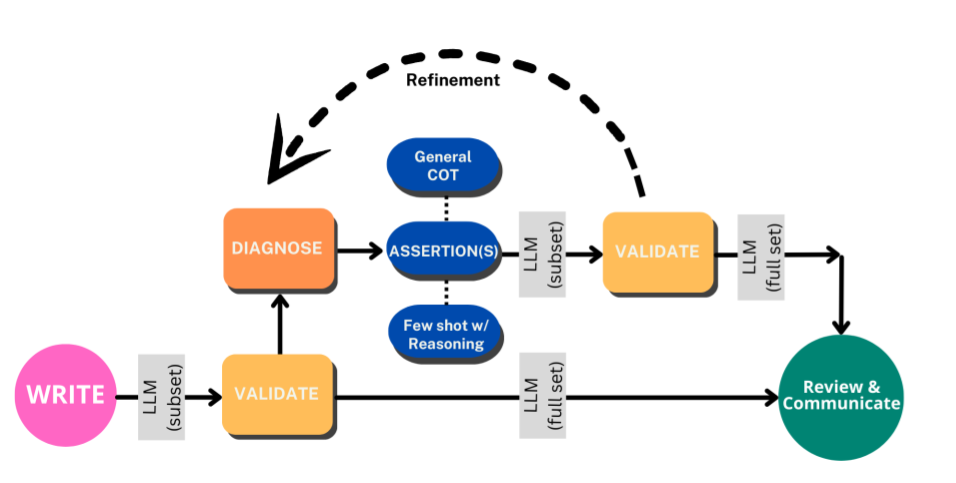}
    \caption{ICL Prompt Engineering Design Process to optimize the accuracy of LLMs in classifying educational data with the use of ICL, COT and AEFL.}
    \label{fig:fig1}
\end{figure}

Our prompt development process, grounded in the ICL Prompt Engineering Design (see Figure \ref{fig:fig1}), begins with drafting an initial few-shot ICL format prompt. This prompt, inputting student responses and outputting CE classifications, undergoes validation testing on a subset (n=27). If benchmarks are met, it progresses to full dataset testing; otherwise, we diagnose misclassifications, realigning LLM outputs with our coding standards through domain-specific CE knowledge integration. Adjustments may involve refining COT processes, FewSHOT learning, or embedding conceptual knowledge assertions. After subset retesting and validation, the optimized prompt is applied to the full dataset (n=135), with iterative refinement ensuring optimal performance. See Appendix \ref{sec:AppendixB} for additional LLM-specific prompt details. 

Our engineering approach encompasses three components: General COT, FewShot with Reasoning Sequence, and assertions Prompting. General COT, embeds sequential instructions with “think time” to initiate the model's reasoning on given tasks \citep{fulford2023chatgpt}. Our General COT prompt follows a seven-step sequence to guide the LLM's task reasoning (see Figure \ref{fig:fig1}). Initially, the model attentively reads the provided <<Question, Response>> (Step 1), laying the foundation for accurate comprehension and subsequent cognitive engagement analysis. Step 2 involves feeding the model CE domain-specific definitions for Passive, Active, and Constructive levels, requiring it to discern the appropriate engagement level based on the initial input. Progressing to Step 3, the model assesses the rationale behind the assigned cognitive engagement label, ensuring it reflects the response's depth and nature. In Step 4, the LLM reevaluates the response to prevent misclassification and assesses if a different CE level is more aligned. Steps 5 and 6 prompt the model to consider ways to enhance the CE level, crucial in the validation and diagnostic phases, particularly when integrating assertions. The final step (Step 7) circles back to the initial input, where the LLM reexamines the cognitive engagement level to verify the accuracy and consistency of its prediction. This structured approach is key in sharpening the model's evaluative and analytical capabilities.

 FewShot with Reasoning, guided by gold standard examples \citep{wang2023investigating, shahriar2023assertion}, includes a four-element structure: <<Question, Response, Label, and Reasoning>>. This method enhances LLM's task-specific learning, incorporating reasoning sequences in the examples. Finally adding assertion Prompts, is crucial for knowledge-building explanations, that are domain-specific insights defined from General COT's outputs on misclassified predictions \citep{shahriar2023assertion}.

\subsection{Experiment Design}
To analyze traditional ML methods (SVM, RF, DT, and ADABoost), we divided our data into training (n=432) and testing sets (n=135), applying default hyperparameters from the Scikit-Learn package  (Pedregosa et al., 2011). See Appendix \ref{sec:AppendixA} for hyperparameters. The dataset comprised two majority classes and one minority class (see Table~\ref{tab:dataset_numbers}. During data preprocessing, we executed text cleaning steps: removing non-alphanumeric/special characters (except periods), new lines, isolated "n" characters, excess spaces, double quotes, and backslashes; converting to lowercase; eliminating stop words; and correcting spelling errors. We transformed the tokenized text using TF-IDF vectorization for ML algorithm suitability. These traditional ML methods served as benchmarks for comparing with LLM prompt results.

\begin{table}[ht]
    \caption{Dataset numbers for Training, Testing and subsets by cognitive level.}
    \centering
    \begin{tabular}{cccc}
        \toprule
        ICAP Level & Training & Testing & Subset \\
        \midrule
        C & 202 & 62 & 10 \\
        A & 203 & 66 & 10 \\
        P & 27 & 7 & 7 \\
        \bottomrule
    \end{tabular}
    \label{tab:dataset_numbers}
\end{table}

In analyzing LLM, we employed GPT-4 through the Colab Python OpenAI API, setting hyperparameters to temperature = 0 and top p= 0.01 for optimal automatic scoring \citep{wang2023investigating}. The data preprocessing mirrored the traditional ML approach but without tokenization or vectorization. We maintained the integrity of student sentences, ensuring capitalized start and appropriate punctuation, mainly periods. The final prompt See Appendix \ref{sec:AppendixB} underwent testing with the same dataset (n=135) used in traditional ML.

In our final experiment, we adopted a subset-based iterative modification approach (n=27) as per the ICL Prompt Design Process \ref{sec:prompt_engineering}. This involved a sensitivity analysis for precise influence measurement of assertions on LLM performance. Each iteration entailed scrutinizing misclassified data, focusing on informal language nuances in text inputs. This qualitative analysis was pivotal for understanding the impact on model accuracy and response. This systematic approach enriched our comprehension of LLM’s interaction with varied prompts and offered insights for enhancing LLM’s performance in processing and interpreting informal language, a significant challenge in educational datasets.

\subsection{Analysis}

In our multiclass dataset analysis, we utilize Precision, Recall, and F1 Score to evaluate the performance of LLMs with assertions versus traditional ML models. These metrics are integral for assessing model efficacy in a multiclass environment. Precision gauges the model’s accuracy in predicting each class, indicating the reliability of its positive predictions. Recall measures the model’s capacity to correctly identify all instances of each class, vital for ensuring comprehensive representation in a multiclass context. The F1 Score, as the harmonic mean of Precision and Recall, offers a balanced evaluation of the model’s overall performance, particularly important in our study to address potential class imbalance. Following \citet{pennebaker2015development}, we emphasize both precision and recall to minimize false positives and negatives, crucial in multiclass datasets. Additionally, we assess the percentage change in F1 score performance to quantify the impact of assertions, using the following formula:

\begin{equation*}
    \text{Percent Increase} = \left(\frac{\text{F1 score of LLM} - \text{F1 score of traditional ML}}{\text{F1 score of traditional ML}}\right) \times 100\%
\end{equation*}

To further this analysis we examined F1 scores. To differentiate between models, we developed a custom metric, inspired by Cohen’s D \citep{cohen2013statistical}. However, unlike the traditional Cohen's D, which uses standardized effect sizes (small at 0.2, medium at 0.5, large at 0.8) based on pooled standard deviation, our metric directly compares raw F1 score differences. This modification suits our data, where standard deviation calculations aren't feasible due to single observations per model. We categorized differences in F1 scores as small (up to 10 points), medium (10 to 30 points), and large (over 30 points).  We defined a function for calculating pairwise differences in scores $m_i$ , $m_j$  M represent any two models, and $s_i$ , $s_j$ are their respective scores. The function:

\begin{center}
$f(m_i, m_j) = s_i - s_j$ is defined as the difference between $s_i$ and $s_j$.
\end{center}

It computes the difference in performance scores between each pair of models. For each combination of models ($m_i$ , $m_j$), the score of model $m_j$ is subtracted from that of model $m_i$. This function calculates the performance difference between each model pair. We then generate a matrix showcasing these differences, allowing for a thorough pairwise comparison of model performances.

To answer RQ 2 and evaluate the ways that the integration of assertions enhance the efficacy of models when addressing the challenges associated with imbalanced textual educational datasets we chose to test on a subset (N=27, P = 10, A = 10, C= 7) as is common in the research to “increase the depth of our analysis, reduce run-time, and decrease cost” \citep[][p. 2]{rodriguez2023prompts}. We chose a sensitivity analysis \citep{akinwande2023understanding} to critically assess the impact or influence of the assertions. We did this qualitatively by adding two steps (Step 5 \& 6 of General COT into the <<General COT>> and interpreting for the <<model outcome>> for recurring themes. Our examination extended to a comparative analysis of the experiments, employing class-wise analysis to measure each experiment against a baseline prompt that did not incorporate assertions.

\section*{Results and Discussion}

\subsection{RQ1: How do the results obtained from LLMs with Prompt Engineering compare to traditional Machine Learning algorithms in handling imbalanced educational data?}

\subsubsection{Performance Metrics}
The summary results in Table \ref{tab:performance_metrics} indicated a varied performance across classes. In the Passive class, the LLM significantly outperformed traditional models, showing a 14.9\% increase over SVM, 6.25\% over RF, 18.0\% over DT, and a notable 23.2\% increase over AdaBoost. Conversely, in the Active class, traditional models (SVM, RF, and DT) surpassed LLM by 11.1\%, while AdaBoost and LLM performances were comparable. The most striking contrast was observed in the Constructive class, where traditional models (SVM, RF, DT, and AdaBoost) failed to effectively identify instances. In contrast, the LLM demonstrated a remarkable improvement with an F1 score of 32, showcasing its superior capability in recognizing elements of the minority class. 

\begin{table}[ht]
    \caption{Summary of Performance Metrics by Cognitive Engagement Level}
    \centering
    \begin{tabular}{c*{9}{c}}
        \toprule
        & \multicolumn{3}{c}{Passive (62)} & \multicolumn{3}{c}{Active (66)} & \multicolumn{3}{c}{Constructive (7)} \\
        \cmidrule(lr){2-4} \cmidrule(lr){5-7} \cmidrule(lr){8-10}
        Model & P & R & F1 & P & R & F1 & P & R & F1 \\
        \midrule
        SVM     & 75 & 73 & 74 & 68 & 77 & 72 & 0  & 0  & 0 \\
        RF      & 70 & 92 & 80 & 80 & 65 & 72 & 0  & 0  & 0 \\
        DT      & 71 & 74 & 72 & 71 & 73 & 72 & 0  & 0  & 0 \\
        ADABoost& 64 & 74 & 69 & 65 & 62 & 64 & 0  & 0  & 0 \\
        LLM     & 83 & 87 & 85 & 78 & 54 & 64 & 21 & 71 & 32 \\
        \bottomrule
    \end{tabular}
    \label{tab:performance_metrics}
\end{table}

\begin{figure}
    \centering
    \includegraphics[width=1\linewidth]{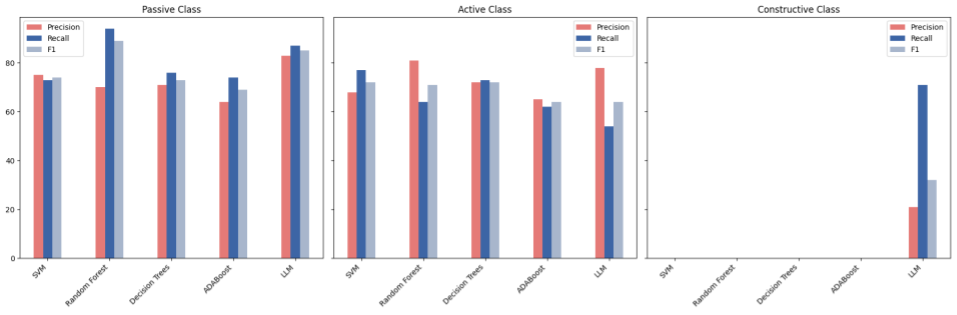}
    \caption{Performance Metrics Summary by Cognitive Engagement Class showing results for each cognitive class.}
    \label{fig:2.performancemetrics}
\end{figure}

These results suggest that while traditional machine learning models like SVM, RF, DT, and AdaBoost may perform comparably or better in majority classes, the LLM exhibits superior capability in dealing with minority class instances, particularly in complex classification tasks like the Constructive class in our dataset (see Figure \ref{fig:2.performancemetrics}). The versatility and adaptability of LLMs in handling imbalanced class distributions highlight their potential in enhancing classification tasks, especially in scenarios where minority classes hold substantial importance. These findings affirm our hypothesis that LLMs, especially when augmented with assertions, offer superior capabilities in classifying cognitive engagement levels from student texts, addressing the core of RQ1.

\subsubsection{Relative Performance}

We see similar results in our custom metric inspired by Cohen's D due to the unique nature of our data, where standard deviation calculations were not applicable, and produced interesting results (see Figure \ref{fig:3.5percentchange}). The LLM with assertions for the passive class demonstrated noteworthy advantages over traditional models in various comparisons which  resonate with the work of researchers \citep{shahriar2023assertion}, who demonstrated the enhanced effectiveness of LLMs in educational settings. 

\begin{figure}
    \centering
    \includegraphics[width=1\linewidth]{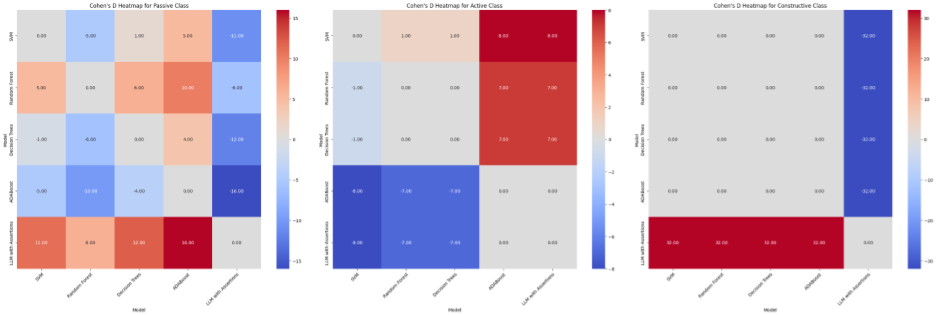}
    \caption{Relative Performance Heatmap by Cognitive Engagement Class}
    \label{fig:3.relativeperformancel}
\end{figure}

Against SVM, the LLM had a significant edge, showing an 11-point advantage in the F1 score, categorized as a 'medium' difference according to our threshold range. This indicates a considerably better performance of the LLM over SVM. When compared to DT, the LLM with assertions again showed a 'medium' difference, outperforming DT by 12 points, underscoring its effectiveness in handling complex classification tasks. In a more striking contrast, the LLM outperformed ADA Boost by 16 points, falling into the 'medium' range and highlighting a substantial performance gap where the LLM was far superior. 

In the Active class, the LLM with assertions exhibited a mixed performance. It showed a close competition with SVM, trailing by just 2 points, which falls into the 'small' difference category, implying a nearly equivalent performance between the two models. However, the LLM outperformed ADABoost by a margin of 8 points, a 'small' difference that nonetheless underscores its relative effectiveness. This suggests that while LLMs offer substantial advantages in many areas, their performance can vary depending on the specific classification task, echoing the findings of \citet{lee2024applying}, who explored the use of LLMs in automatic scoring systems. Against RF and DT, the LLM had a slight disadvantage, trailing by 7 and 3 points respectively, suggesting that in certain scenarios, traditional models may have a slight edge over the LLM. 

The Constructive class results were particularly striking. The LLM with assertions demonstrated a pronounced superiority in this category. It dramatically outperformed all traditional models (SVM, RF, DT, and ADABoost), each of which failed to identify instances within the Constructive class effectively, as indicated by their zero scores. The LLM achieved an F1 score of 32, which not only establishes a 'large' difference according to our threshold but also highlights the LLMs exceptional capability in handling minority classes or complex classification tasks where traditional models fall short. It points to the LLMs' superior ability to handle imbalanced datasets, a common challenge in educational data analysis, as illustrated by the work of researchers like \cite{zeng2023challenge}, who evaluated the cognitive and reasoning abilities of LLMs.

\subsection{RQ2: In what ways does the integration of assertions enhance the efficacy of models when addressing the challenges associated with imbalanced textual educational datasets?}

Our analysis aimed to augment Active class metrics and foster a more equitable model across cognitive classes. Throughout the course of ten experiments, including the baseline, the implementation of assertions, particularly those delineated in <<General COT>> (Steps 5 \& 6, see Appendix \ref{sec:AppendixB}), was pivotal in surfacing two primary themes post the initial experiment: textual ambiguity and contextual comprehension challenges.

For text ambiguity, the baseline experiment revealed the model's propensity to misconstrue the depth of student engagement. Instances where contributions appeared analytical but merely constituted a superficial application of known concepts underscored this issue. By systematically applying the assertions detailed in the Methodology, we observed significant improvements in model performance, particularly within the Active and Constructive classes. 

With regard to Unusual language, the model's interpretation of speculative language (e.g., "I think," "possibly," "I believe") as indicative of reflective or analytical thought. Such expressions, particularly when conveying opinions that superficially suggested deeper analysis, were erroneously classified as constructive engagement.

Initially, our approach to integrating assertions was exploratory but became more systematic by the third experiment. For example, between experiments two through four, certain responses intended as "Constructive" were incorrectly classified as "Active":

\noindent Misclassified Example 1: 
\begin{quote}
  Question: Why do you think the model learned a large negative weight for this feature?
Student response: “I think the model learned a negative weight for this feature because the model categorized the reviews as negative and categorized the surprisingly negative features as negative too since that was the whole sentiment of the review.”  
\end{quote}

\noindent Misclassified Example 2: 
\begin{quote}
    Question: Why do you think the model learned a large positive weight for this feature?
Student response: “I feel like it had to do with the words and how much they were used whenever there was a positive review it would contain more than one good word to go along with it”
\end{quote}

By incorporating the assertion <<Do label the statement as Constructive when they form a hypothesis about why the model learned a weight for a certain feature>>, these responses were accurately predicted as constructive, enhancing the Constructive class with precision and recall metrics—specifically, a recall increase of 6.33\% and an F1-score improvement of 4.30\%.

\begin{figure}
    \centering
    \includegraphics[width=1\linewidth]{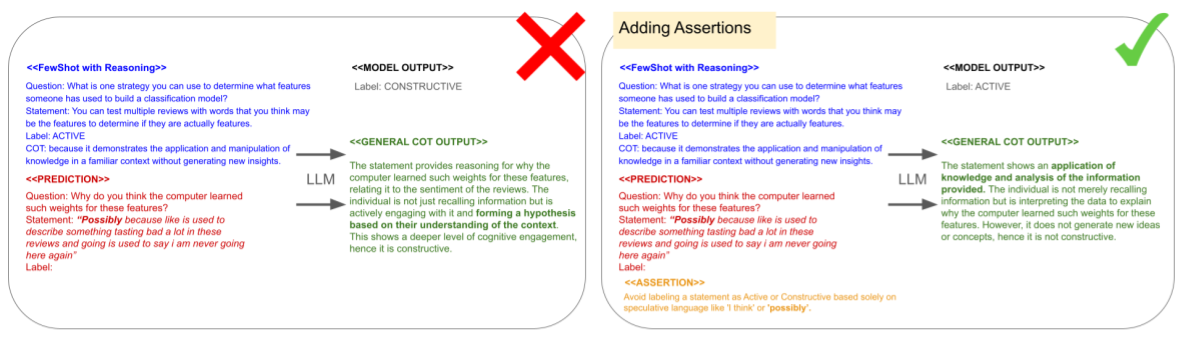}
    \caption{Left image does not include a targeted assertion while the one on the right does and improves the model output to correctly predict the students cognitive level of their text.}
    \label{fig:3.4targetedassert}
\end{figure}

Moreover, addressing the misuse of speculative language through the assertion <<Avoid labeling a statement as Active or Constructive based solely on speculative language like 'I think' or 'possibly’>> (see Figure \ref{fig:3.4targetedassert}) led to an increase in precision for the Active class by 15.96\% and an F1-score increase by 6.08\%. This adjustment resulted in the most balanced model performance observed, despite a slight decrease in recall for the Active class by 2.34\%. Further attempts to amplify Active class metrics by refining definitions in <<General COT>> and enhancing reasoning in <<FewShot with reasoning>> revealed that, while assertions impacted model performance, their effect varied across classes and metrics.  

\begin{figure}
    \centering
    \includegraphics[width=1\linewidth]{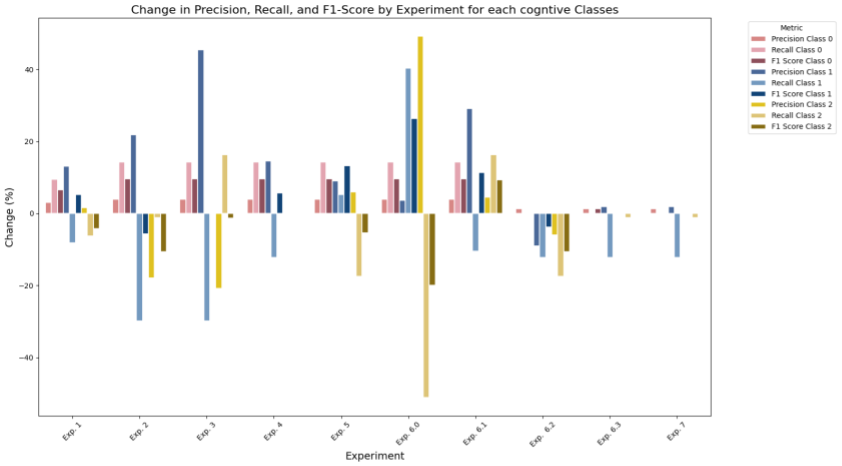}
    \caption{Percentage Change in Metrics for Each Class Across Experiments}
    \label{fig:3.5percentchange}
\end{figure}

Notably, Experiment 6.1 (see Figure \ref{fig:3.5percentchange}) emerged as particularly effective, showcasing the significance of tailored assertions in reducing misclassifications linked to textual ambiguity and unusual language use, thereby contributing to a more balanced and accurate model.

These findings highlight the nuanced role of assertions in enhancing model efficacy against the backdrop of imbalanced educational datasets. By meticulously integrating assertions to counter specific challenges—textual ambiguity and unusual language—the experiments demonstrated a discernible improvement in model precision and balance, particularly within the Active and Constructive classes. This strategic approach underscores the potential of assertions to mitigate inherent dataset imbalances, ultimately contributing to the development of more nuanced and effective educational models.

To further understand our model we compared accuracy of models to the baseline where Experiment 5 marked an 8.96\% improvement but Experiment 6.1 stood out with the highest increase in accuracy, at 11.94\% from the baseline. This improvement primarily addresses the challenges identified in RQ2, demonstrating the significant role of assertions in resolving errors related to context understanding and lexical ambiguities. The Active and Constructive classes, associated with focused attention and deeper reasoning, respectively, pose classification challenges due to their subtleties and contextual dependencies \citep{chi2014icap}. These classes often require inferring cognitive engagement levels from implicit cues and context, making their distinctions less explicit within student responses.

\section{Limitations}
While our study sheds light on the potential of LLMs and AEFL in addressing imbalanced datasets, it also highlights the need for caution in interpreting these findings without consideration of the broader methodological and technological landscape. Firstly, our reliance on specific LLM techniques and AEFL might not capture the full spectrum of potential solutions available within the rapidly evolving field of machine learning. The specific parameters and configurations employed in our LLM applications \citep{shahriar2023assertion, wei2022chain, zeng2023challenge}, while effective in this context, might not be universally applicable or optimal across different datasets or learning tasks. While our study provides valuable insights, it echoes the concerns raised by \citet{radwan2017improving} and \citet{yun2011effective} regarding the challenges of imbalanced datasets in education and the limitations of traditional ML approaches. 

Furthermore, our study's focus on a AI High School ELA course dataset \citep{zeng2023challenge}, while providing a rich source of cognitive engagement data, also presents a limitation in terms of diversity and representativeness. The linguistic and cognitive patterns inherent in this specific educational setting may not fully encapsulate the variety of cognitive engagement manifestations across different age groups, subjects, or educational methodologies. This limitation underscores the importance of extending research efforts to encompass a wider range of educational contexts, to ensure the findings' applicability and robustness, as indicated by \citet{fredricks2004school} and \citet{blumenfeld2006motivation}.

Additionally, while LLMs and AEFL present innovative approaches to overcoming the challenges of imbalanced datasets, they also introduce new complexities and considerations \citep{shahriar2023assertion, wei2022chain}. The computational demands and resource requirements of these technologies, coupled with the need for specialized expertise to implement and interpret their outputs, may pose barriers to widespread adoption and application in educational research and practice. The dynamic nature of LLM development also means that the models and techniques used today may rapidly evolve, necessitating continuous updates and adaptations to maintain their effectiveness and relevance.

Lastly, the ethical implications of applying LLMs in educational settings, particularly concerning data privacy, security, and the potential for bias in model training and outcomes, warrant careful consideration \citep{zeng2023challenge}. As LLMs become more integrated into educational research and practice, it is crucial to develop and adhere to ethical guidelines that prioritize the well-being and rights of students and educators.

These limitations highlight the need for ongoing research and dialogue within the educational and machine learning communities. By addressing these challenges and exploring the vast potential of LLMs and AEFL, we can advance our understanding of cognitive engagement and enhance educational outcomes in diverse and inclusive ways.

\section{Conclusion and Future Studies}
Our study makes significant contributions to the evolving landscape of cognitive engagement (CE) research, building upon the foundational work of seminal researchers like \citet{craik1972levels}, \citet{appleton2006measuring}, and \citet{fredricks2004school}. We leveraged the capabilities of Large Language Models (LLMs) and Assertion Enhanced Few-Shot Learning (AEFL), marking a notable advancement in the domain of CE. This approach pays homage to the pioneering efforts that have shaped our understanding of CE while extending these concepts through the integration of cutting-edge LLM technologies.

By adeptly navigating the challenges posed by imbalanced datasets and accurately classifying cognitive engagement levels, this study underscores the potential of LLMs to refine our measurement and analysis of CE, setting a new benchmark for educational research. The integration of AEFL enhances contextual comprehension, improving model accuracy and balance, as highlighted by \citet{shahriar2023assertion}. Experiment 6.1 further illustrates the value of tailored assertions in reducing misclassifications linked to textual ambiguities, offering novel insights into AEFL's effectiveness in managing class-imbalanced data.

The promising outcomes of this research suggest that LLMs hold significant potential for future educational studies, particularly in complex data analysis tasks. These findings encourage the exploration of LLMs' full capabilities in educational settings, advocating for a paradigm shift towards more sophisticated and nuanced approaches to data analysis. Moreover, the integration of AEFL points to a nuanced method of enhancing model performance, especially in the context of imbalanced textual educational datasets.

Given the multifaceted nature of cognitive engagement and the challenges associated with its measurement, there is a compelling need for further research. Future studies should aim to refine and expand the application of LLMs and AEFL across a broader spectrum of educational contexts. Additionally, exploring additional theoretical frameworks and models could yield deeper insights into cognitive engagement, thereby contributing to the enhancement of educational outcomes. This call for further research not only reflects the complex landscape of CE but also highlights the endless possibilities that LLM technologies and innovative methodologies like AEFL present for advancing our understanding and practices within the educational domain.

\bibliographystyle{unsrtnat}
\bibliography{references}  %%% Uncomment this line and comment out the ``thebibliography'' section below to use the external .bib file (using bibtex) .

\begin{thebibliography}{49}
\providecommand{\natexlab}[1]{#1}
\providecommand{\url}[1]{\texttt{#1}}
\expandafter\ifx\csname urlstyle\endcsname\relax
  \providecommand{\doi}[1]{doi: #1}\else
  \providecommand{\doi}{doi: \begingroup \urlstyle{rm}\Url}\fi

\bibitem[Karimah and Hasegawa(2022)]{karimah2022automatic}
Shofiyati~Nur Karimah and Shinobu Hasegawa.
\newblock Automatic engagement estimation in smart education/learning settings: a systematic review of engagement definitions, datasets, and methods.
\newblock \emph{Smart Learning Environments}, 9\penalty0 (1):\penalty0 31, 2022.

\bibitem[Radwan and Cataltepe(2017)]{radwan2017improving}
Akram~M Radwan and Zehra Cataltepe.
\newblock Improving performance prediction on education data with noise and class imbalance.
\newblock \emph{Intelligent Automation \& Soft Computing}, pages 1--8, 2017.

\bibitem[Yun et~al.(2011)Yun, Nan, Da, and Bing]{yun2011effective}
ZHAI Yun, Ma~Nan, Ruan Da, and AN~Bing.
\newblock An effective over-sampling method for imbalanced data sets classification.
\newblock \emph{Chinese Journal of Electronics}, 20\penalty0 (3):\penalty0 489--494, 2011.

\bibitem[Fredricks et~al.(2004)Fredricks, Blumenfeld, and Paris]{fredricks2004school}
Jennifer~A Fredricks, Phyllis~C Blumenfeld, and Alison~H Paris.
\newblock School engagement: Potential of the concept, state of the evidence.
\newblock \emph{Review of educational research}, 74\penalty0 (1):\penalty0 59--109, 2004.

\bibitem[Blumenfeld et~al.(2006)Blumenfeld, Kempler, and Krajcik]{blumenfeld2006motivation}
Phyllis~C Blumenfeld, Toni~M Kempler, and Joseph~S Krajcik.
\newblock \emph{Motivation and cognitive engagement in learning environments}.
\newblock na, 2006.

\bibitem[Corno and Mandinach(1983)]{corno1983role}
Lyn Corno and Ellen~B Mandinach.
\newblock The role of cognitive engagement in classroom learning and motivation.
\newblock \emph{Educational psychologist}, 18\penalty0 (2):\penalty0 88--108, 1983.

\bibitem[Pintrich(2000)]{pintrich2000role}
Paul~R Pintrich.
\newblock The role of goal orientation in self-regulated learning.
\newblock In \emph{Handbook of self-regulation}, pages 451--502. Elsevier, 2000.

\bibitem[Schunk et~al.(2014)Schunk, Pintrich, and Meece]{schunk2014motivation}
Dale~H Schunk, Paul~R Pintrich, and Judith~L Meece.
\newblock Motivation in education: Theory, research, and applications.
\newblock \emph{(No Title)}, 2014.

\bibitem[Shahriar et~al.(2023)Shahriar, Matsuda, and Ramos]{shahriar2023assertion}
Tasmia Shahriar, Noboru Matsuda, and Kelly Ramos.
\newblock Assertion enhanced few-shot learning: Instructive technique for large language models to generate educational explanations.
\newblock \emph{arXiv preprint arXiv:2312.03122}, 2023.

\bibitem[Brown et~al.(2020)Brown, Mann, Ryder, Subbiah, Kaplan, Dhariwal, Neelakantan, Shyam, Sastry, Askell, et~al.]{brown2020language}
Tom Brown, Benjamin Mann, Nick Ryder, Melanie Subbiah, Jared~D Kaplan, Prafulla Dhariwal, Arvind Neelakantan, Pranav Shyam, Girish Sastry, Amanda Askell, et~al.
\newblock Language models are few-shot learners.
\newblock \emph{Advances in neural information processing systems}, 33:\penalty0 1877--1901, 2020.

\bibitem[Wei et~al.(2022{\natexlab{a}})Wei, Tay, Bommasani, Raffel, Zoph, Borgeaud, Yogatama, Bosma, Zhou, Metzler, et~al.]{wei2022emergent}
Jason Wei, Yi~Tay, Rishi Bommasani, Colin Raffel, Barret Zoph, Sebastian Borgeaud, Dani Yogatama, Maarten Bosma, Denny Zhou, Donald Metzler, et~al.
\newblock Emergent abilities of large language models.
\newblock \emph{arXiv preprint arXiv:2206.07682}, 2022{\natexlab{a}}.

\bibitem[Zeng et~al.(2023)Zeng, Chen, Jiang, and Jia]{zeng2023challenge}
Zhongshen Zeng, Pengguang Chen, Haiyun Jiang, and Jiaya Jia.
\newblock Challenge llms to reason about reasoning: A benchmark to unveil cognitive depth in llms.
\newblock \emph{arXiv preprint arXiv:2312.17080}, 2023.

\bibitem[Wu(2021)]{wu2021learning}
Jiun-Yu Wu.
\newblock Learning analytics on structured and unstructured heterogeneous data sources: Perspectives from procrastination, help-seeking, and machine-learning defined cognitive engagement.
\newblock \emph{Computers \& Education}, 163:\penalty0 104066, 2021.

\bibitem[Wei et~al.(2022{\natexlab{b}})Wei, Wang, Schuurmans, Bosma, Xia, Chi, Le, Zhou, et~al.]{wei2022chain}
Jason Wei, Xuezhi Wang, Dale Schuurmans, Maarten Bosma, Fei Xia, Ed~Chi, Quoc~V Le, Denny Zhou, et~al.
\newblock Chain-of-thought prompting elicits reasoning in large language models.
\newblock \emph{Advances in neural information processing systems}, 35:\penalty0 24824--24837, 2022{\natexlab{b}}.

\bibitem[Lee et~al.(2024)Lee, Latif, Wu, Liu, and Zhai]{lee2024applying}
Gyeong-Geon Lee, Ehsan Latif, Xuansheng Wu, Ninghao Liu, and Xiaoming Zhai.
\newblock Applying large language models and chain-of-thought for automatic scoring.
\newblock \emph{Computers and Education: Artificial Intelligence}, page 100213, 2024.

\bibitem[Craik and Lockhart(1972)]{craik1972levels}
Fergus~IM Craik and Robert~S Lockhart.
\newblock Levels of processing: A framework for memory research.
\newblock \emph{Journal of verbal learning and verbal behavior}, 11\penalty0 (6):\penalty0 671--684, 1972.

\bibitem[Appleton et~al.(2006)Appleton, Christenson, Kim, and Reschly]{appleton2006measuring}
James~J Appleton, Sandra~L Christenson, Dongjin Kim, and Amy~L Reschly.
\newblock Measuring cognitive and psychological engagement: Validation of the student engagement instrument.
\newblock \emph{Journal of school psychology}, 44\penalty0 (5):\penalty0 427--445, 2006.

\bibitem[Chi and Wylie(2014)]{chi2014icap}
Michelene~TH Chi and Ruth Wylie.
\newblock The icap framework: Linking cognitive engagement to active learning outcomes.
\newblock \emph{Educational psychologist}, 49\penalty0 (4):\penalty0 219--243, 2014.

\bibitem[Anderson and Krathwohl(2001)]{anderson2001taxonomy}
Lorin~W Anderson and David~R Krathwohl.
\newblock \emph{A taxonomy for learning, teaching, and assessing: A revision of Bloom's taxonomy of educational objectives: complete edition}.
\newblock Addison Wesley Longman, Inc., 2001.

\bibitem[Bloom et~al.(1956)]{bloom1956taxonomy}
Benjamin~S Bloom et~al.
\newblock Taxonomy of educational objectives: The classification of educational goals, by a committee of college and university examiners.
\newblock \emph{Handbook 1: Cognitive domain}, 1956.

\bibitem[Chase et~al.(2019)Chase, Marks, Malkiewich, and Connolly]{chase2019teacher}
Catherine~C Chase, Jenna Marks, Laura~J Malkiewich, and Helena Connolly.
\newblock How teacher talk guidance during invention activities shapes students’ cognitive engagement and transfer.
\newblock \emph{International Journal of STEM Education}, 6:\penalty0 1--22, 2019.

\bibitem[Hsiao et~al.(2022)Hsiao, Chen, Chen, and Lin]{hsiao2022developing}
Jo-Chi Hsiao, Ssu-Kuang Chen, Wei Chen, and Sunny~SJ Lin.
\newblock Developing a plugged-in class observation protocol in high-school blended stem classes: Student engagement, teacher behaviors and student-teacher interaction patterns.
\newblock \emph{Computers \& Education}, 178:\penalty0 104403, 2022.

\bibitem[Wang et~al.(2016)Wang, Wen, and Ros{\'e}]{wang2016towards}
Xu~Wang, Miaomiao Wen, and Carolyn~P Ros{\'e}.
\newblock Towards triggering higher-order thinking behaviors in moocs.
\newblock In \emph{Proceedings of the Sixth International Conference on Learning Analytics \& Knowledge}, pages 398--407, 2016.

\bibitem[McCoach et~al.(2013)McCoach, Gable, Madura, McCoach, Gable, and Madura]{mccoach2013defining}
D~Betsy McCoach, Robert~K Gable, John~P Madura, D~Betsy McCoach, Robert~K Gable, and John~P Madura.
\newblock Defining, measuring, and scaling affective constructs.
\newblock \emph{Instrument development in the affective domain: School and corporate applications}, pages 33--90, 2013.

\bibitem[Greene et~al.(2004)Greene, Miller, Crowson, Duke, and Akey]{greene2004predicting}
Barbara~A Greene, Raymond~B Miller, H~Michael Crowson, Bryan~L Duke, and Kristine~L Akey.
\newblock Predicting high school students' cognitive engagement and achievement: Contributions of classroom perceptions and motivation.
\newblock \emph{Contemporary educational psychology}, 29\penalty0 (4):\penalty0 462--482, 2004.

\bibitem[Smiley and Anderson(2011)]{smiley2011measuring}
Whitney Smiley and Robin Anderson.
\newblock Measuring students' cognitive engagement on assessment tests: A confirmatory factor analysis of the short form of the cognitive engagement scale.
\newblock \emph{Research \& Practice in Assessment}, 6:\penalty0 17--28, 2011.

\bibitem[Lee and Anderson(1993)]{lee1993task}
Okhee Lee and Charles~W Anderson.
\newblock Task engagement and conceptual change in middle school science classrooms.
\newblock \emph{American educational research journal}, 30\penalty0 (3):\penalty0 585--610, 1993.

\bibitem[Helme and Clarke(2001)]{helme2001identifying}
Sue Helme and David Clarke.
\newblock Identifying cognitive engagement in the mathematics classroom.
\newblock \emph{Mathematics Education Research Journal}, 13\penalty0 (2):\penalty0 133--153, 2001.

\bibitem[Wigfield et~al.(2008)Wigfield, Guthrie, Perencevich, Taboada, Klauda, McRae, and Barbosa]{wigfield2008role}
Allan Wigfield, John~T Guthrie, Kathleen~C Perencevich, Ana Taboada, Susan~Lutz Klauda, Angela McRae, and Pedro Barbosa.
\newblock Role of reading engagement in mediating effects of reading comprehension instruction on reading outcomes.
\newblock \emph{Psychology in the Schools}, 45\penalty0 (5):\penalty0 432--445, 2008.

\bibitem[Xie et~al.(2019)Xie, Heddy, and Greene]{xie2019affordances}
Kui Xie, Benjamin~C Heddy, and Barbara~A Greene.
\newblock Affordances of using mobile technology to support experience-sampling method in examining college students' engagement.
\newblock \emph{Computers \& Education}, 128:\penalty0 183--198, 2019.

\bibitem[D'Mello et~al.(2017)D'Mello, Dieterle, and Duckworth]{d2017advanced}
Sidney D'Mello, Ed~Dieterle, and Angela Duckworth.
\newblock Advanced, analytic, automated (aaa) measurement of engagement during learning.
\newblock \emph{Educational psychologist}, 52\penalty0 (2):\penalty0 104--123, 2017.

\bibitem[Bernacki et~al.(2012)Bernacki, Byrnes, and Cromley]{bernacki2012effects}
Matthew~L Bernacki, James~P Byrnes, and Jennifer~G Cromley.
\newblock The effects of achievement goals and self-regulated learning behaviors on reading comprehension in technology-enhanced learning environments.
\newblock \emph{Contemporary Educational Psychology}, 37\penalty0 (2):\penalty0 148--161, 2012.

\bibitem[Ireland and Henderson(2014)]{ireland2014language}
Molly~E Ireland and Marlone~D Henderson.
\newblock Language style matching, engagement, and impasse in negotiations.
\newblock \emph{Negotiation and conflict management research}, 7\penalty0 (1):\penalty0 1--16, 2014.

\bibitem[Lee and Kinzie(2012)]{lee2012teacher}
Youngju Lee and Mable~B Kinzie.
\newblock Teacher question and student response with regard to cognition and language use.
\newblock \emph{Instructional science}, 40:\penalty0 857--874, 2012.

\bibitem[Chawla(2010)]{chawla2010data}
Nitesh~V Chawla.
\newblock Data mining for imbalanced datasets: An overview.
\newblock \emph{Data mining and knowledge discovery handbook}, pages 875--886, 2010.

\bibitem[Fern{\'a}ndez et~al.(2018)Fern{\'a}ndez, Garc{\'\i}a, Galar, Prati, Krawczyk, and Herrera]{fernandez2018learning}
Alberto Fern{\'a}ndez, Salvador Garc{\'\i}a, Mikel Galar, Ronaldo~C Prati, Bartosz Krawczyk, and Francisco Herrera.
\newblock \emph{Learning from imbalanced data sets}, volume~10.
\newblock Springer, 2018.

\bibitem[Kulkarni et~al.(2020)Kulkarni, Chong, and Batarseh]{kulkarni2020foundations}
Ajay Kulkarni, Deri Chong, and Feras~A Batarseh.
\newblock Foundations of data imbalance and solutions for a data democracy.
\newblock In \emph{Data democracy}, pages 83--106. Elsevier, 2020.

\bibitem[Japkowicz and Stephen(2002)]{japkowicz2002class}
Nathalie Japkowicz and Shaju Stephen.
\newblock The class imbalance problem: A systematic study.
\newblock \emph{Intelligent data analysis}, 6\penalty0 (5):\penalty0 429--449, 2002.

\bibitem[Bruce et~al.(2020)Bruce, Bruce, and Gedeck]{bruce2020practical}
Peter Bruce, Andrew Bruce, and Peter Gedeck.
\newblock \emph{Practical statistics for data scientists: 50+ essential concepts using R and Python}.
\newblock O'Reilly Media, 2020.

\bibitem[Lema{\~A}{\v{Z}}tre et~al.(2017)Lema{\~A}{\v{Z}}tre, Nogueira, and Aridas]{lemaavztre2017imbalanced}
Guillaume Lema{\~A}{\v{Z}}tre, Fernando Nogueira, and Christos~K Aridas.
\newblock Imbalanced-learn: A python toolbox to tackle the curse of imbalanced datasets in machine learning.
\newblock \emph{Journal of machine learning research}, 18\penalty0 (17):\penalty0 1--5, 2017.

\bibitem[Savelka et~al.(2023)Savelka, Denny, Liffiton, and Sheese]{savelka2023efficient}
Jaromir Savelka, Paul Denny, Mark Liffiton, and Brad Sheese.
\newblock Efficient classification of student help requests in programming courses using large language models.
\newblock \emph{arXiv preprint arXiv:2310.20105}, 2023.

\bibitem[Cui et~al.(2023)Cui, Zhang, Li, Lopez, Das, Malin, and Kumar]{cui2023divide}
Wendi Cui, Jiaxin Zhang, Zhuohang Li, Damien Lopez, Kamalika Das, Bradley Malin, and Sricharan Kumar.
\newblock A divide-conquer-reasoning approach to consistency evaluation and improvement in blackbox large language models.
\newblock In \emph{Socially Responsible Language Modelling Research}, 2023.

\bibitem[Chao et~al.(2022)Chao, Finzer, Ros{\'e}, Jiang, Yoder, Fiacco, Murray, Tatar, and Wiedemann]{chao2022storyq}
Jie Chao, Bill Finzer, Carolyn~P Ros{\'e}, Shiyan Jiang, Michael Yoder, James Fiacco, Chas Murray, Cansu Tatar, and Kenia Wiedemann.
\newblock Storyq: a web-based machine learning and text mining tool for k-12 students.
\newblock In \emph{Proceedings of the 53rd ACM Technical Symposium on Computer Science Education V. 2}, pages 1178--1178, 2022.

\bibitem[Fulford and Ng(2023)]{fulford2023chatgpt}
Andrew Ng~Isa Fulford and A~Ng.
\newblock Chatgpt prompt engineering for developers. deeplearning. ai, 2023.

\bibitem[Wang et~al.(2023)Wang, Wang, Xu, Geng, Zhang, Tao, Rudzicz, Mercer, and Jiang]{wang2023investigating}
Xindi Wang, Yufei Wang, Can Xu, Xiubo Geng, Bowen Zhang, Chongyang Tao, Frank Rudzicz, Robert~E Mercer, and Daxin Jiang.
\newblock Investigating the learning behaviour of in-context learning: a comparison with supervised learning.
\newblock \emph{arXiv preprint arXiv:2307.15411}, 2023.

\bibitem[Pennebaker et~al.(2015)Pennebaker, Boyd, Jordan, and Blackburn]{pennebaker2015development}
James~W Pennebaker, Ryan~L Boyd, Kayla Jordan, and Kate Blackburn.
\newblock The development and psychometric properties of liwc2015.
\newblock 2015.

\bibitem[Cohen(2013)]{cohen2013statistical}
Jacob Cohen.
\newblock \emph{Statistical power analysis for the behavioral sciences}.
\newblock Routledge, 2013.

\bibitem[Rodriguez et~al.(2023)Rodriguez, Dearstyne, and Cleland-Huang]{rodriguez2023prompts}
Alberto~D Rodriguez, Katherine~R Dearstyne, and Jane Cleland-Huang.
\newblock Prompts matter: Insights and strategies for prompt engineering in automated software traceability.
\newblock In \emph{2023 IEEE 31st International Requirements Engineering Conference Workshops (REW)}, pages 455--464. IEEE, 2023.

\bibitem[Akinwande et~al.(2023)Akinwande, Jiang, Sam, and Kolter]{akinwande2023understanding}
Victor Akinwande, Yiding Jiang, Dylan Sam, and J~Zico Kolter.
\newblock Understanding prompt engineering may not require rethinking generalization.
\newblock \emph{arXiv preprint arXiv:2310.03957}, 2023.

\end{thebibliography}

%%% Uncomment this section and comment out the \bibliography{references} line above to use inline references.
% \begin{thebibliography}{1}

% 	\bibitem{kour2014real}
% 	George Kour and Raid Saabne.
% 	\newblock Real-time segmentation of on-line handwritten arabic script.
% 	\newblock In {\em Frontiers in Handwriting Recognition (ICFHR), 2014 14th
% 			International Conference on}, pages 417--422. IEEE, 2014.

% 	\bibitem{kour2014fast}
% 	George Kour and Raid Saabne.
% 	\newblock Fast classification of handwritten on-line arabic characters.
% 	\newblock In {\em Soft Computing and Pattern Recognition (SoCPaR), 2014 6th
% 			International Conference of}, pages 312--318. IEEE, 2014.

% 	\bibitem{hadash2018estimate}
% 	Guy Hadash, Einat Kermany, Boaz Carmeli, Ofer Lavi, George Kour, and Alon
% 	Jacovi.
% 	\newblock Estimate and replace: A novel approach to integrating deep neural
% 	networks with existing applications.
% 	\newblock {\em arXiv preprint arXiv:1804.09028}, 2018.

% \end{thebibliography}

%ACKNOWLEDGMENTS are optional
\section{Acknowledgments}
This material is based upon work supported by the National Science Foundation under Grant No. DRL-1949110. Any opinions, findings, conclusions, or recommendations expressed in this material are those of the author(s) and do not necessarily reflect the views of the National Science Foundation.

\newpage

\begin{appendices}
%Appendix 
\section{Appendix A}\label{sec:AppendixA}

(\begin{table}[h!]
    \centering
    \caption{Support Vector Machine (SVM), Random Forest (RF), Decision Trees (DT), ADABoost}
    \begin{tabular}{c c c}
     \toprule
     \textbf{ML Classifier} & \textbf{Parameter} & \textbf{Values} \\
     \midrule
      SVM & Kernel & `linear', `sigmoid', `poly', `rbf' \\
      & C & 0.1, 1, 10, 100, 1000 \\
      & Gamma & 1, 0.1, 0.01, 0.001, 0.0001 \\
      RF & Number of Estimators & 100 \\
      & Criterion & `Gini' \\
      & Max Depth & none \\
      & Min Samples Leaf & 1 \\
      & Min Samples Split & 2 \\
      & Max Features & `Auto' \\
      & Bootstrap & True \\
      DT & Criterion & `Gini' \\
      & Max Depth & none \\
      & Min Samples Leaf & 1 \\
      & Min Samples Split & 2 \\
      ADABoost & Number of Estimators & 50 \\
      & Learning Rate & 1.0 \\
      & Algorithm & `SAMME.R' \\
      & Base Estimator & DecisionTreeClassifier(max\_depth=1) \\
      \bottomrule
    \end{tabular}
    \label{tab:model_parameters}
\end{table})

\section{Appendix B}\label{sec:AppendixB}

\noindent A. Few Shot with reasoning + General COT (step by step)  < Black, \textcolor{teal}{Green}, \textcolor{red}{Red}, \textcolor{blue}{Blue}>

\noindent B. Few Shot with reasoning +  General COT (step by step) + Assertion (do and don’t)<Black, \textcolor{teal}{Green}, \textcolor{red}{Red}, \textcolor{blue}{Blue}, \textcolor{orange}{Orange}>

\begin{verbatim}
-----------Prompt Starts Here—------------------------------------
\end{verbatim}

\noindent Your task is to identify the label of the statement delimited by triple backticks
\color{teal}

\noindent Read the instructions below:

\noindent Step 1: Read the question and statement attentively to understand the context and the nature of the statement provided.

\noindent Step 2: Determine the initial cognitive engagement level of the statement using the definitions of the provided cognitive engagement labels - passive, active, and constructive.

\noindent 1. Passive engagement: a statement  is classified as "Passive" when the individual is only receiving information without interacting with it or adding anything to it. Passive engagement typically involves listening, reading, or receiving information without actively processing, manipulating, or reflecting upon it.

\noindent 2.  Active engagement: a statement is classified as "Active" when the response involves applying knowledge, analyzing information, or manipulating information but not generating new ideas or concepts.

\noindent 3. Constructive engagement: a statement is classified as "Constructive" if it reflects reasoning, justification, or thoughtful consideration based on prior knowledge.

\noindent Step 3: Assess why it corresponds to the label you placed it in. Consider the extent to which it demonstrates recall of basic information (passive), application of learned knowledge to slightly different contexts (active), or a deeper level of analysis and synthesis of various concepts (constructive).

\noindent Step 4: Critically evaluate whether the statement could potentially belong to other labels. Examine the nuances of the statement to see if there are elements that might indicate a higher or lower level of cognitive engagement.

\noindent Step 5: To upgrade the statement to a higher engagement level, propose alterations that would make it align with the criteria for the "Active" category. This could involve adding details that show the application of learned knowledge to familiar yet slightly different contexts, or demonstrating problem-solving based on previous experiences.

\noindent Step 6: Explore how the statement can be restructured to meet the criteria of the "Constructive" engagement category. Consider adding elements that showcase deeper analysis, critical evaluation, or synthesis of multiple concepts to create a more nuanced and thoughtful response.
Step 7: Finally, revisit the question and statement to evaluate the original cognitive engagement level making sure the prediction of cognitive engagement is accurate.

\color{red}

\noindent Based on your understanding of cognitive engagement and the labeled examples provided, determine the level of engagement for the unlabeled text provided. 
\begin{verbatim}
    ```
\end{verbatim}
\noindent Question: Why do people write reviews?

\noindent Statement: People write reviews to express their feelings on a certain thing to condemn a praise a business, franchise, movie, or book.

\noindent Label: <Generate label>

\noindent Chain-of-thought: <Generate the chain-of-thought>
\begin{verbatim}
    ```
\end{verbatim}

\color{blue}

\noindent Use the following examples delimited by triple quotes to understand which label the statement belongs to.
\begin{verbatim}
    '''
\end{verbatim}
\noindent Question: What features do you think are indicators of positive reviews?

\noindent Statement: Words like love, excellent, greatest, amazing, enjoy, awesome, best.

\noindent Label: Passive

\noindent Reasoning: because it is a direct response that involves recalling or listing words without further analysis or interaction.

\noindent Question: What is one strategy you (as a human) can use to determine if a review is positive or negative?

\noindent Statement: I can tell if the person liked something or not.
\noindent Label: Passive
\noindent Reasoning: because it does not specify any strategies or reflection to distinguish between positive and negative sentiments.

\noindent Question: When you click on the row, the feature in this review will be highlighted in the feature graph (like the one you have seen in the Light On Light Off activity).Which feature do you think is it?

\noindent Statement: Because it's associated with positivity.

\noindent Label: Passive

\noindent Reasoning: because it is simple information without reflection without delving into specific details, analysis, or reflection.

\noindent Question: What is one strategy you can use to determine what features someone has used to build a classification model?

\noindent Statement: I can use major words that people say in reviews first. Words like 'love,' 'hate,' 'bad,' 'delicious,' and more.

\noindent Label: Passive

\noindent Reasoning: because it only has recall words and delve into any analysis, reflection, or application.

\noindent Question: What is one strategy you can use to determine what features someone has used to build a classification model?

\noindent Statement: You can look at the data set and find words that really stand out to you or words that have a strong emotional connotation. You can also check the graphand the probability in terms of the features being used or how strongly they correlate with the result.

\noindent Label: Active

\noindent Reasoning: because it summarizes and organizes the information in a broad manner

\noindent Question: What is one strategy you (as a human) can use to determine if a review is positive or negative?
Statement: One strategy that you can use to determine if a review is positive or negative is looking at diction, which is word choice, and how the words are being used.

\noindent Label: Active

\noindent Reasoning: because it details a method of analyzing the word choice in reviews, demonstrating the application of acquired knowledge to assess sentiments.

\noindent Question: When you click on the row, the feature in this review will be highlighted in the feature graph (like the one you have seen in the Light On Light Off activity). Which feature do you think is it?

\noindent Statement: Love is the most defining word in this review, if it were changed to 'hate' it would have a completely different meaning

\noindent Label: Active

\noindent Reasoning: because it demonstrates the application and analysis of knowledge in a familiar context but does not generate new ideas.

\noindent Question: What is one strategy you can use to determine what features someone has used to build a classification model?

\noindent Statement: You can test multiple reviews with words that you think may be the features to determine if they are actually features.

\noindent Label: Active

\noindent Reasoning: because it demonstrates the application and manipulation of knowledge in a familiar context without generating new insights.

\noindent Question: Why do people write reviews?

\noindent Statement: To share their experience of a certain product or service so that they can either warn or recommend it to people. Sharing experiences is important so that way others who have not experienced it can know what they are getting in to.

\noindent Label: Constructive

\noindent Reasoning: because it provides an understanding and reasoning of the broader context and implications why sharing experience is important.

\noindent Question: If none of the 10 features are present in your review, try again with another review. If some of the 10 features are in your review, examine both your review and the feature graph. What do you think these features are?

\noindent Statement: I think these features are key words and numbers. Like the example used the word 'love' which implies a positive reply. The numbers also because if you say 1 out of 10 that's bad but if you say 10 out of 10 that's good.

\noindent Label: Constructive
\noindent Reasoning: because it provides interpretation and application to generate insights about the potential features in reviews.

\noindent Question: What is one strategy you (as a human) can use to determine if a review is positive or negative?

\noindent Statement: If I am having a conversation with somebody it will be easy to detect if the review is good or bad by word choice and their tone. If they wrote it, I will be able to see key words that point in either a positive or negative direction.

\noindent Label: Constructive

\noindent Reasoning: because it demonstrates a depth of reasoning and reflection of how to determine if a review is positive or negative.

\noindent Question: What kinds of reviews can make our world a better place?

\noindent Statement: Some reviews that can make the world a better place is if it's a review about a foreign country then it can give some insight into what is happening within that country. Or even here in the United States, it can share what's happening within their state and let the rest of the world know.

\noindent Label: Constructive

\noindent Reasoning: because it provides reflection, thoughtful consideration and reasoning about the societal value and potential impact of reviews in fostering global understanding and awareness.
\begin{verbatim}
    '''
\end{verbatim}

\color{orange}

\noindent A few facts about identifying the cognitive engagement level that you must assert while determining the level of engagement for the unlabeled text provided:

\noindent - Do label the statement as Constructive if they are forming an opinion about its usefulness, and providing reasoning for their opinion.

\noindent - Do label the statement as Constructive when the statement provides their interpretation and reasoning to the question.

\noindent - Do label the statement as Constructive when they form a hypothesis about why the model learned a  weight for a certain feature.

\noindent - Do label the statement as Constructive when the statement shows active engagement with the information.

\noindent - Avoid labeling a statement as Active or Constructive based 
solely on speculative language like 'I think' or 'possibly’.

\end{appendices}
\end{document}